\documentclass[onecolumn,draftclsnofoot,12pt]{IEEEtran}
\usepackage{graphicx}
\usepackage{color}
\usepackage{amsfonts}
\usepackage{amsmath,amsthm}
\newtheorem{mydef}{Theorem}
\usepackage{amssymb}
\usepackage{cite}

\newcommand{\abc}[1]{{\color{blue}#1}}
\renewcommand{\abc}[1]{#1}
\newcommand{\abm}[1]{{\color{red}[Abhishek Comment:] #1}}
\renewcommand{\abm}[1]{}
\newcommand{\source}{\mathcal{S}}
\newcommand{\tx}[3]{\mathcal{T}_{#1,#3}}
\newcommand{\sink}{\mathcal{K}}
\newcommand{\rx}[3]{\mathcal{R}_{#1,#3}}
\newcommand{\x}{\mathbf{x}}
\newcommand{\pair}{pair~}
\newcommand{\link}{link~}
\newcommand{\erf}[1]{\mathsf{erf}\left(#1\right)}
\newcommand{\erfc}[1]{\mathsf{erfc}\left(#1\right)}
\newcommand{\spp}{\Phi}
\newcommand{\se}{\mathrm{se}}

\ifCLASSINFOpdf
  
\else
  
\fi

\begin{document}

\title{On Hybrid MoSK-CSK Modulation based Molecular Communication: Error Rate Performance Analysis using Stochastic Geometry}

\author{\IEEEauthorblockN{Nithin V. Sabu$^\dagger$, Neeraj Varshney$^\ddagger$ and Abhishek K. Gupta$^\dagger$}
\IEEEauthorblockA{$^\dagger$Department of Electrical Engineering, Indian Institute of Technology Kanpur, Kanpur, India 208016.\\
$^\ddagger$Wireless Networks Division, National Institute of Standards and Technology, Maryland, USA 20899.\\
Email: nithinvs@iitk.ac.in, neerajv@ieee.org and gkrabhi@iitk.ac.in}}

\pagenumbering{gobble}
\maketitle

\begin{abstract}
Data transmission rate in molecular communication systems can be improved by using multiple transmitters and receivers. In molecular multiple-input multiple-output (MIMO) systems which use only single type of molecules, the performance at the destination is limited by inter-symbol interference (ISI), inter-link interference (ILI) and multi-user interference (MUI). This work proposes a new hybrid modulation for a system with multiple transmitters and receivers which uses different types of molecules to eliminate ILI. Further, to enhance the data rate of the proposed system under ISI and MUI, $\textbf{M}$-ary CSK  modulation scheme is used between each transmitter-receiver pair. In this paper, the random locations of transmitters present in the three dimensional (3-D) space are modeled as homogeneous Poisson point process (HPPP). Using stochastic geometry tools,  analytical expression is derived for the probability of symbol error for the aforementioned scenario. Finally, the performance of the proposed system is compared using the different existing modulation schemes such as on-off keying (OOK), binary concentration shift keying (BCSK) and quadruple concentration shift keying (QCSK) to develop several important insights.
\end{abstract}

\IEEEpeerreviewmaketitle

\section{Introduction}

Molecular Communication is a promising solution in situations where conventional communication methods are impractical and inefficient, for example, communication in the saline environment, inside tunnels, pipelines and between nano-machines \cite{farsad2016comprehensive}. Examples of molecular communication systems include bacterial based communication, systems involving microtubules and motor proteins and  systems implementing molecular communication via diffusion (MCvD). In the recent literature, MCvD systems are studied in detail compared to other communication methods. A link implementing MCvD between source-sink pair consists of the following process blocks: encoding, transmission, propagation, reception and decoding \cite{nakano2013molecular}. For a molecular communication system involving bio-nanomachines (BNMs), the transmitter located at the source BNM translates the message symbols into information molecules. These molecules are then released to the propagation medium and move to the receiver via diffusion.  The receptor structures present in the receiver located at the sink BNM captures these information molecules for detection process and remove them from the environment. The received molecules are used to decode the transmitted information.

In molecular communication, many modulation schemes including concentration shift keying (CSK), molecular shift keying (MoSK), pulse position modulation, isomer based ratio shift keying and depleted MoSK (D-MoSK) have been proposed \cite{kuran2011modulation,llatser2013detection,kim2013novel,kabir2015d}. Modulation schemes can help in improving the performance of systems, but, for applications requiring high reliability and high speed data transmission, the performance of SISO systems is not satisfactory. Therefore, the use of multiple transmitters  at the source and multiple receivers at the sink BNM can be leveraged to solve the aforementioned problem leading to MIMO molecular systems  \cite{meng2012mimo}. Various techniques like transmit diversity, selection combining, maximal-ratio combining and decision fusion can be used to improve the performance of MIMO molecular communication systems. 
The work in \cite{meng2012mimo} numerically demonstrated that the spatial diversity reduces the error probability whereas the spatial multiplexing improves the data rate. However, the MIMO systems may suffer from inter-link interference (ILI).  ILI arises since receivers cannot identify whether the received molecules are emitted from the intended transmitter or the interfering transmitters from the same source. A MIMO system with point transmitters and spherical absorbing receivers which includes ISI due to residual molecules in the environment from previous symbol transmission and ILI due to other transmitters in the same system emitting the same type of molecules was studied in  \cite{koo2016molecular}. In \cite{koo2016molecular}, the authors demonstrated that the use of MIMO with spatial multiplexing increases the data rate by 1.7 times compared to the SISO counterpart. In a molecular communication system, the fraction of observed molecules at the receiver represent the channel coefficient or channel impulse response (CIR). For a molecular MIMO system with absorbing receivers, the analytical expressions for  CIR  is not available in the past literature.  Therefore, the work in \cite{lee2017machine} has tried to model a molecular MIMO channel by an artificial neural network (ANN). Further, a trained ANN is used to acquire channel response for performance evaluation of spatially coded MIMO systems in \cite{damrath2017spatial}. 

A molecular source-sink communication \pair can coexist with other communication pairs in the same medium leading to a multi-pair molecular communication system (MP-MCS). This system may suffer from multi-user interference (MUI). The location of the BNMs in 3D spaces can be modeled using Poisson point process (PPP). For example, the spatial distribution of bacterial colonies inside cheese was proven to fit the Poisson process \cite{Jeanson1493}. A SISO system with multiple BNMs with their location modeled in 3-D space as a spatial homogeneous Poisson point process (HPPP) was studied in \cite{pierobon2014statistical}. Using stochastic geometry, a general model for collective signal strength in a large-scale SISO system with and without molecule degradation was discussed in \cite{deng2017analyzing}.  Also, the bit error probability for a passive receiver and a fully absorbing receiver was derived in \cite{deng2017analyzing}.

A multi-\pair MIMO molecular system suffers from all three types of interferences: ISI, ILI and MUI \cite{koo2016molecular,damrath2017spatial}. Due to ILI, in MIMO systems with point transmitters and fully-absorbing spherical receivers, the CIR is obtained using simulation based approaches \cite{koo2016molecular,damrath2017spatial} owing to lack of analytical expression. Also, for multi-\pair MIMO molecular systems with multiple transmitters and receivers,  there is no work in literature which models the location of interferers in 3-D space as random.

In this paper, we consider a molecular communication system having multiple transmitters at the source and receivers at the sink BNM  in a 3-D medium and propose a hybrid modulation scheme combining MoSK and CSK. 
The random locations of these source BNMs are modeled as uniform PPP in the 3-D space outside the tagged receiver volume. In the proposed hybrid MoSK-CSK modulation based system, each transmitter-receiver \link in a source-sink \pair uses $M$-ary CSK modulation with different transmitter-receiver links using different type of molecules. Thus, the modulation scheme consists of MoSK along with $M$-ary CSK to form hybrid modulation. The use of different type of molecules across links completely avoids inter-link interference at any receiver of a sink BNM since the molecules from undesired transmitters of the paired source do not interfere. 
Due to the absence of ILI,  the analytical expression for channel response can be applied to the system instead of just the numerical results via simulation-based approaches. Also, the molecules of same type from other interfering sources arriving at the receiver reduces, and hence MUI also reduces, compared to that of MIMO systems using the single type of molecules for all links. Using stochastic geometry, we develop an analytical framework to study this system and derive the analytical expression for probability of symbol error for various $M$-ary CSK modulation schemes. 
\\
\textit{Notations:} Boldface lowercase letters ({\em e.g. $\x$}) represent coordinates/vectors. $\erf{\cdot},\ \erfc{\cdot}$ and  $\mathbb{E}[\cdot ]$ denote error function, complementary error function and expectation operator, respectively.

\begin{figure}
\begin{center}
\includegraphics[scale=0.54]{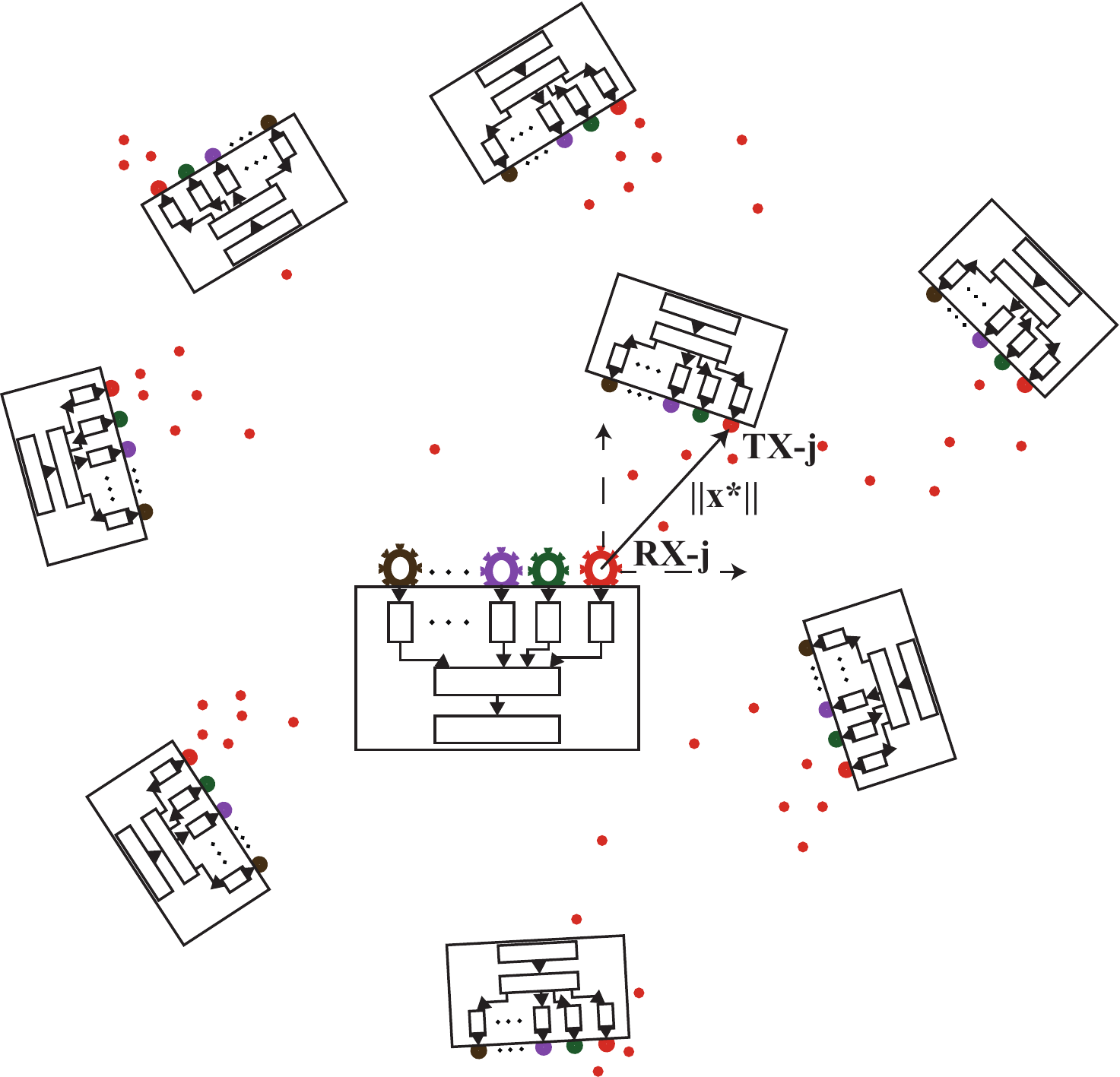}\\
\end{center}
\caption{Schematic diagram of a tagged receiver at origin and the source BNMs distributed as uniform PPP in a 3-D environment.} 
\label{fig:1}
\end{figure}

\begin{figure*}
\begin{center}
\includegraphics[scale=0.6]{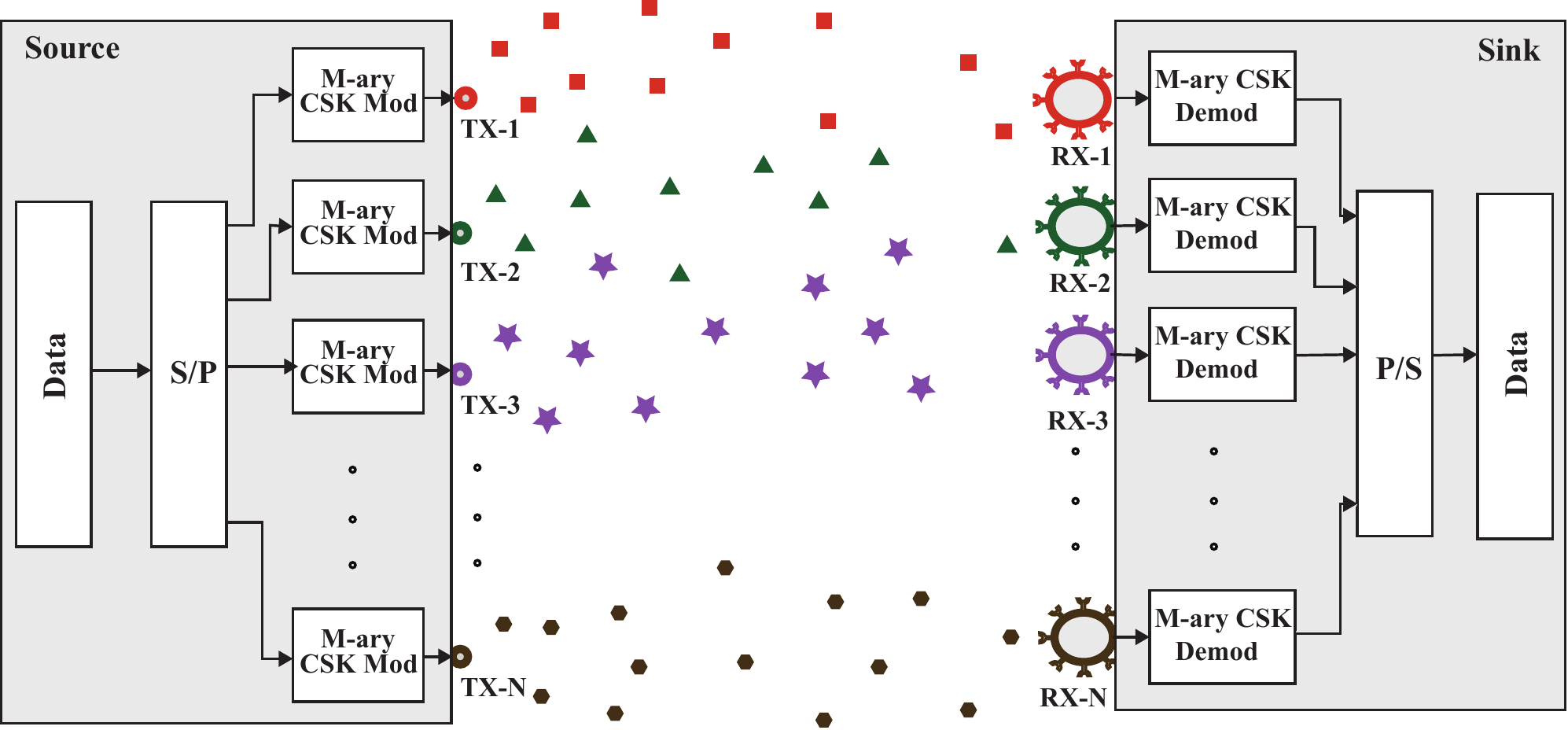}\\
\end{center}
\caption{Schematic diagram of a source-sink pair, where different colours represent transmitter-receiver links use different types of molecules.}  \label{fig:2}
\end{figure*}
\section{System Model}
In this paper, we consider a molecular communication system with multiple sources and one sink as shown in Fig. \ref{fig:1}, where only one source randomly communicates to a sink and others act as interfering sources in a 3-D homogeneous pure-diffusive fluid medium.
\subsection{Transmitter-Receiver Links}
The schematic diagram of a single source-sink pair is shown in Fig. \ref{fig:2}. Each  source-sink pair consists of a source BNM having $N$ point transmitters attached to it and a sink BNM with $N$ fully absorbing spherical receivers. The $n$th receiver has radius $r_n$ and volume $V_n$.  
A typical transmitter-receiver link between $n$th transmitter and $n$th receiver uses $M$-ary CSK modulation and use molecules of type-$n$ different from the molecules used by the other transmitter-receiver links. As shown in Fig. \ref{fig:2}, the source first converts the serial information bits to $N$ parallel bit sequences of length $\log_2{M}$ at each time slot, and then each point transmitter emits molecules to the propagation medium according to $M$-ary CSK modulation. Therefore, the bit rate $(R_b)$ of this system is $R_b=N\log_2(M)$ per symbol period.  At the sink, $N$ spherical receivers with receptors which bind to only one type of molecule count the absorbed molecules and after demodulation, the obtained bit sequence is finally converted to serial form. Similar to several existing works \cite{koo2016molecular,damrath2017spatial,deng2017analyzing}, the transmitter-receiver couples are assumed to be perfectly synchronized, and the spherical absorbing receiver can count all of the absorbed molecules. The source and sink are assumed to be entirely transparent to the signal molecule diffusion, except for absorption by the corresponding receiver structures. Further, as mentioned earlier, this work considers that each of the receivers can detect only the type of information molecule which is emitted by its coupled transmitter. This assumption is justified since the number of molecule types that can be identified by a BNM is constrained due to its limited size and limited complexity\cite{suda2005exploratory}. Moreover, the advantage of using different type of information molecules for different transmitter-receiver link is the elimination of ILI which is prevalent in systems with multiple transmitters and receivers which use single type of information molecules. The separation between the spherical receivers in the proposed system also has no impact on the performance due to the absence of ILI.
\subsection{Source-Sink Pairs}
 The locations of the centroid of all sources are modeled as HPPP in a three dimensional (3-D) space  of $\mathbb{R}^3$ with density $\lambda$. Let $\spp_j=\{\x_{ji},i\in\mathbb{N}\}$ be the point process representing locations of $j$th point transmitter $\tx{\source_i}{}{\x_{ji}}$ of all source $\source_i$. The $i$th sink is denoted by $\sink_i$ and is paired with $\source_i$. Independent displacement of points in PPP according to some probability distribution will also be PPP according to displacement theorem \cite{baccelli2010stochastic}. Therefore, $\spp_j$ is a PPP.  Let us consider a typical pair $\source_0-\sink_0$  where a $j$th receiver $\rx{\sink_0}{}{\mathbf{0}}$ of sink $0$ is considered at the origin. For this tagged receiver, the $\spp_j$ can be modeled as PPP outside the  receiver $(\mathbb{R}^3\setminus V_j)$ with intensity $\lambda$. The desired transmitter $\tx{\source_0}{}{\x_{j0}}$ of the desired source $\source_0$ is a part of the PPP $\spp_j$, as allowed by Slivnyak theorem \cite{baccelli2010stochastic} without affecting the results. For analysis, we consider that there are multiple sources and only a single sink in the 3-D space since the analytical formula for CIR is not available  in the current existing literature for a system with multiple fully-absorbing spherical receivers which absorbs the same type of information molecules.
\subsection{Molecular Signal with Degradation}
For a point source located at a distance $r$ away from the center of an spherical receiver of radius $r_n$,the hitting rate of molecules at the receiver is given as \cite{schulten2000lectures},
\begin{equation}
q(t{'}|r)=\frac{r_n}{r}\frac{r-r_n}{\sqrt{4\pi D {t{'}}^3}}\exp{\left \{  -\frac{(r-r_n)^2}{4Dt{'}}\right \},}
\end{equation}
where $D$ denotes the diffusion coefficient, which depends on the characteristics of the molecule and the fluid environment. The detection performance can be improved by introducing adequate amount of molecular degradation \cite{heren2015effect}. The molecule should reach the receiver boundary before its degradation and it should not reside in the environment more than symbol time in order to avoid ISI.  The rate of degradation of molecule $(\mu_d)$ depends on the half-life $(\Lambda_{1/2})$ of molecule undergoing degradation, i.e., $\mu_d=\ln(2)/\Lambda_{1/2}$.

The fraction of non-degraded molecules absorbed by the spherical receiver within an arbitrary time $t$ since transmission time can be obtained as \cite{heren2015effect},
\begin{align}
f(\mu_d,t\mid r)=&\int_{0}^{t}q(t^{'}|r)e^{-\mu_d t^{'}}dt^{'}\nonumber\\
=&\frac12 \frac{r_n}{r}\exp\left( -\sqrt{\frac{\mu_d}{D}}(r-r_n)\right)
\times\left[
	\exp\left(
		2\sqrt{\frac{\mu_d}{D}}(r{-}r_n)
		\right)
	\erfc{
		\frac{r{-}r_n}{\sqrt{4Dt}} +\sqrt{\mu_d t} 
	}
 \right. \nonumber\\
 &\left.
+\erfc{\frac{r-r_n}{\sqrt{4Dt}} -\sqrt{\mu_d t} }
\right]\label{eq:fmh}.
\end{align}
The above equation shows that the fraction of molecules absorbed at the spherical receiver reduces with increase in $\mu_d$ \cite{heren2015effect}. For the scenario with no molecular degradation, {\em i.e.},  $\mu_d=0$, the above expression simplifies as,
\begin{equation}
f(0,t\mid r)=\frac{r_n}{r}\erfc{ \frac{r-r_n}{\sqrt{4Dt}}}.
\end{equation}
\subsection{Channel Model}

As stated earlier, each of the transmitter-receiver link uses $M$-ary CSK modulation for communication. At time instant $k$, the \abc{$j$th} ($1\leq j\leq N $) transmitter $\tx{\source_i}{j}{\x_{ji}}$  of the \abc{$i$th} source machine \abc{$\source_i$} randomly located at \abc{$\x_{ji}=\x$} emits $u_{\x}^{\abc{j}}[k]= Q_m, \ 0\leq m \leq M-1$  molecules at the beginning of symbol period $T_s$ corresponding to the message symbol $s_{\textbf{x}}^j[k]=S_m$. For example, in OOK modulation, the $j$th transmitter emits either no molecules ($Q_0=0$) or $Q_1$ number of molecules corresponding to bit 0 or 1. 
\

\abc{Let us consider a typical $j$th receiver of pair $\source_0-\sink_0$, located at the origin. We assume a} discrete time channel model with channel memory of length $L$, where $h_{\x}^{j}[l]$ denotes the channel impulse response at $l$th time instant for the channel between \abc{ the typical receiver and the $j$th transmitter of the $i$th source $\source_i$} which is randomly located at $\x$ according to uniform PPP. The CIR at the $l$th time instant, which is the fraction of molecules absorbed between $lT_s$ and $(l+1)T_s$ can be obtained from \eqref{eq:fmh} and is given as, 
\abc{\begin{align}
h_{\x}^{j}[l]=f(\mu_d,(l+1)T_s\mid \|\x\|)-f(\mu_d,lT_s\mid \|\x\|).
\end{align}
}

Let the $j$th transmitter of the desired pair source $\source_0$ be located at $\x^*$. Considering the hitting of information molecules on the receiver as success events, the number of molecules  \abc{that were transmitted from $\tx{\source_0}{i}{\x}$ and  detected} by the $j$th spherical receiver at $k$th time instant follows Binomial distribution with parameters $(u_{\textbf{x}}^j[ k-l ],h_{\textbf{x}}^{j} [ l ])$. 
We approximate it to Poisson distribution \abc{assuming} the number of information molecules $u_{\x}^j[ k-l ]$ is large and the hitting probability $h_{\textbf{x}}^{j} [ l ]$ is small. Note that the sum of independent Poisson random variables follows Poisson distribution. Thus, the total number of molecules received at the $j$th receiver in $k$ th time instant $y_\mathbf{0}^j[k]$  follows Poisson distribution with parameter \abc{$\sum\limits_{\x\in \spp_j}\sum\limits_{l=0}^{L} h_{\x }^{j} [ l ]u_{\x}^j[ k-l ]$}. That is,
\begin{align}
\abc{y_\mathbf{0}^j[k]}\sim&\mathcal{P}\left(\overbrace{h_{\textbf{x}^* }^{j}\left [ 0 \right ]u_{\textbf{x}^*}^j\left [ k \right ]}^{\text{Desired}}+\overbrace{\sum_{l=1}^{L}h_{\textbf{x}^*}^{j}\left [ l \right ]u_{\textbf{x}^*}^j\left [ k-l \right ]}^{\text{ISI}}\right.\left.+\overbrace{\sum_{\textbf{x}\in \Phi_j\setminus\left \{ \textbf{x}^* \right \}}\sum_{l=0}^{L}h_{\textbf{x}}^{j}\left [ l \right ]u_{\textbf{x}}^j\left [ k-l \right]}^{\text{MUI}}\right), \label{Nee1}
\end{align}
where $\mathcal{P}(.)$ represents Poisson distribution. The number of molecules received at the tagged receiver located at the origin is the sum of molecules corresponding to the desired symbol, previous symbol molecules which causes ISI, and molecules of the same type from other interfering transmitters which result in MUI. 

\section{Probability of Symbol Error Analysis}
For simplicity, we consider that all the $j$th transmitters of all sources of multiple pair system are sending the same symbols \abc{$s^j[k]$} with probability of sending $S_m$ as $P_{S_m}$ at any time instant $k$. The sink BNM uses fixed threshold detection in which at every symbol period the counted number of molecules are compared with the predetermined threshold values $(\tau_0,\tau_1,...,\tau_{M})$ and is decoded as $S_m$ if the received molecules count is between $\tau_m$ and $\tau_{m+1}$. Let the decoded symbol at the $k$th time instant be \abc{$\hat{s}^j[k]$} and $P^j_{m,\se}[k]=P^j\left (\hat{s}^j[k]\neq S_m\mid s^j[k]=S_m ,s^j[k-L-1:k-1]\right )$ at the $k$th time instant be the probability of event that the sink fails to decode $\hat{s}^j[k]=S_m$ corresponding to the tagged receiver $j$ given that the transmitted symbol is $s^j[k]=S_m$ and the previous symbols $s^j[k-L-1:k-1]$. The probability of symbol error $P_{\se}^j[k]$ during time instant $k$ at tagged receiver conditioned on $L$ previous symbols  is given by
\begin{align}
P_{\abc{\se}}^j[k]= \sum_{m=0}^{M-1}P_{S_m} \underbrace{P^j\left (\hat{s}^j[k]{\neq} S_m{\mid} s^j[k]{=}S_m, s^j[k{-}L{-}1{:}k{-}1]\right )}_{\triangleq P^j_{m,\se}[k]}.\!\!\label{eq:e1}
\end{align}

\subsection{System with One Source and One Sink}
To solve the expression for $P^j_{m,\se}[k]$ in \eqref{eq:e1}, we first consider a system with only one source-sink pair which implies lack of MUI. In this scenario, the number of molecules received at the tagged \abc{receiver} at the $k$th time slot follows Poisson distribution with parameter 
$\sum_{l=0}^{L} h^{j}_{\x^*} [ l ]\abc{u^j_{\x^*}}[ k-l ]$. Here the only source of interference is due to the ISI of the desired transmitter located at $\x^*$. The $m$th symbol is not decoded as $S_m$ when the absorbed number of molecules at the $k$th time slot is not in the limit of lower and upper threshold values $\tau_{m}$ and $\tau_{m+1}$.  Then,
\begin{align}
P^j_{m,\se}[k]=&1-\sum_{n=\tau_{m}}^{\tau_{m+1}}\frac{1}{n!}\exp
	\left(
		-\displaystyle\sum_{l=0}^{L} h^{j}_{\x^*} [ l ]\abc{u^j_{\x^*}}[ k-l ]
	\right)
\times\left(\displaystyle\sum_{l=0}^{L} h^{j}_{\textbf{x}^*} [ l ]u^j_{\x^*}[ k-l ]\right)^n.\label{eq:e2}
\end{align}
\begin{figure}
	\begin{center}
		\includegraphics[scale=.405]{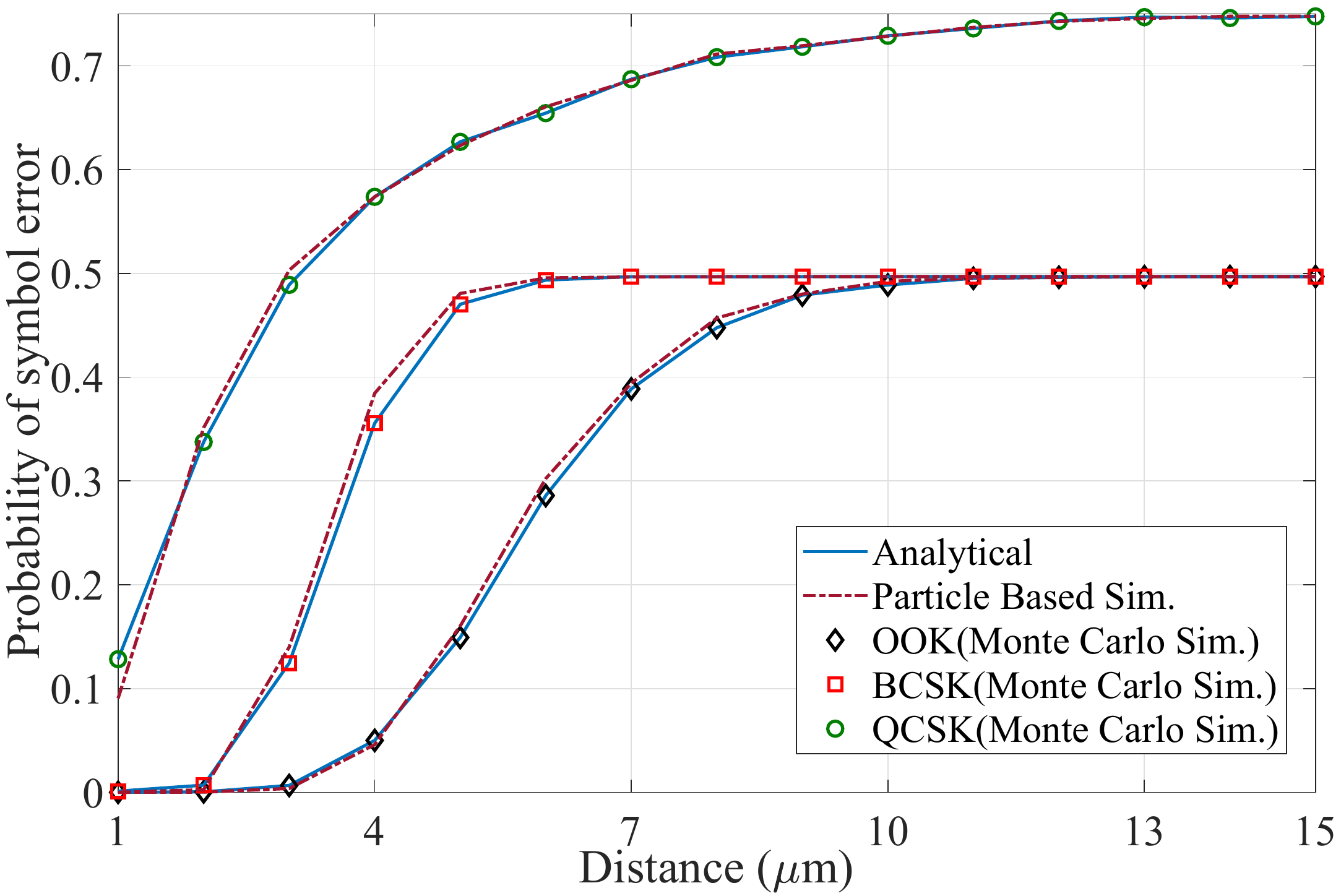}
	\end{center}
	\caption{Symbol error probability versus distance between point transmitter and spherical receiver $(r-r_j)$ in a system with single link. The probability of symbol error increases with an increase in distance between point transmitter and spherical receiver. OOK modulation performs better than others.}
	\label{fig:3}
\end{figure}

\subsection{System with Multiple Sources and One Sink}
We now consider a system with multiple sources which includes both ISI and MUI. 
\abc{Let $z^j(k, \|\textbf{x}\|)$ be the number of molecules absorbed by tagged receiver that were emitted by the $j$th transmitter located  at $\x$, {\em i.e.,}}
\begin{equation}
z^j(k, \|\x\|)=\abc{\sum_{l=0}^{L}h_{\x}^{j}\left [ l \right ]u_{\x}^{j}\left [ k-l \right]}
\end{equation}
and let $Z^j(k)$ be the total number of received molecules from all $j$th transmitters, {\em i.e.,}
\begin{equation}
Z^j(k)=\sum_{\x\in \spp_j}z^j(k,\|\x\|).
\end{equation}
Then, the expression for $P^j_{m,\se}[k]$ can be solved as,
\begin{align}
&P^j_{m,\se}[k]=1{-}\mathbb{E}_{\spp_j}\left [\sum_{n=\tau_{m}}^{\tau_{m+1}} \frac1{n!}{\exp{(-Z^j(k))}(Z^j(k))^n}\right].\!\label{eq:e3}
\end{align}
\abc{Note that, 
\begin{equation}
\exp\left ( -Z^j(k) \right)(- Z^j(k))^n=\left.\frac{\mathrm{d}^n\exp\left ( - \rho Z^j(k) \right)}{\mathrm{d}\rho^n}\right\vert_{\rho=1}.
\end{equation}
Now, from \eqref{eq:e3} we get
\begin{align}
P^j_{m,\se}[k]=&1-\mathbb{E}_{\spp_j}\left [\sum_{n=\tau_{m}}^{\tau_{m+1}} \frac{\exp{( -Z^j(k))}(-Z^j(k))^n}{(-1)^nn!}\right]\nonumber\\
=&1-\sum_{n=\tau_{m}}^{\tau_{m+1}}\frac{1}{(-1)^nn!}\left.\frac{\mathrm{d}^n \mathcal{L}_{Z^j(k)}(\rho)}{\mathrm{d}\rho^n}\right\vert_{\rho=1}.\label{eq:e4}
\end{align}
Here, $ \mathcal{L}_{Z^j(k)}(\rho)$ is the Laplace functional of $Z^j(k)$ which can be obtained as
\begin{align}
\mathcal{L}_{Z^j(k)}(\rho)&=\mathbb{E}\left [ \exp\left( -\rho\sum_{\x\in \spp_j}z^j(k, \|\x\|) \right)\right ]\nonumber\\
&=\exp\left ( -4\pi \lambda\int_{r_j}^{\infty}(1-\exp( -\rho z^j(k, r) )) r^2\mathrm{d}r\right ).\label{eq:e5}
\end{align}}
Using the Laplace functional \eqref{eq:e5} and Bell polynomial version of Faa di Bruno's formula, \cite[eq.(2.2)]{rio},
\begin{align}
\left.\frac{d^n \mathcal{L}_{Z^j(k)}(\rho)}{d\rho^n}\right\vert_{\rho=1}=\exp\left \{ -4\pi \lambda\int_{r_j}^{\infty}(1-\exp\{ - z^j(k, r)  \}) r^2dr\right \}\times(-1)^n\times \sum_{k=0}^n\mathfrak{B}_{n,k} (\mathbf{p}(k,\lambda)),\label{eq:e6}
\end{align}
where $\mathbf{p}(k,\lambda)=[p_{1}(k,\lambda),p_{2}(k,\lambda),...,p_{n-k+1}(k,\lambda)]$ and \[p_i(k,\lambda)= 4\pi \lambda\int_{r_j}^{\infty}\exp\{ -z^j(k, r)\} (z^j(k , r) )^i r^2dr.\] The quantity $\mathfrak{B}_{n,k} (\mathbf{p}(k,\lambda))$ in \eqref{eq:e6} is the incomplete exponential Bell polynomial which can be represented as,
\begin{align}
\mathfrak{B}_{n,k} (\mathbf{p}(k,\lambda))&=\sum\frac{n!}{j_1!j_2!...j_{n-k+1}!} \prod_{v=1}^{n-k+1}\left(\frac{p_v(k,\lambda)}{v!}\right)^{j_v}.\label{eq:e7}
\end{align}
In \eqref{eq:e7} the sum is taken over all non-negative integers $j_1,j_2,...,j_{n-k+1}$ such that $j_1+j_2+...+j_{n-k+1}=k$ and $1j_1+2j_2+...+(n-k+1)j_{n-k+1}=n$. Substituting \eqref{eq:e6} in \eqref{eq:e4} , we get the following theorem.

\abc{
\begin{mydef}
{The probability of symbol error for $m$th symbol in a multi-pair molecular system with hybrid MoSK-CSK modulation is given as
\begin{align}
P^j_{m,\se}=&P^j\left (\hat{s}^j[k]\neq S_m\mid s^j[k]=S_m ,s^j[k-L-1:k-1]\right )\nonumber\\
=&1-\exp\left \{ -4\pi \lambda\int_{r_j}^{\infty}(1-\exp\{ -z(k, r)  \})r^2dr\right \} \times \sum_{n=\tau_{m}}^{\tau_{m+1}}\frac{1}{n!}\mathfrak{B}_n (\mathbf{p}(k,\lambda)),\label{eq:e8}
\end{align}
where $\mathfrak{B}_n(\mathbf{p}(k,\lambda)) =\sum_{k=0}^n\mathfrak{B}_{n,k}(\mathbf{p}(k,\lambda))$. $\mathfrak{B}_{0,0}(\mathbf{p}(k,\lambda))=1$ and $\sum_{k=1}^n\mathfrak{B}_{n,k}(\mathbf{p}(k,\lambda))$ is the $n$th complete exponential Bell polynomial.}}
\end{mydef}
\begin{figure}
	\begin{center}
		\includegraphics[scale=0.39]{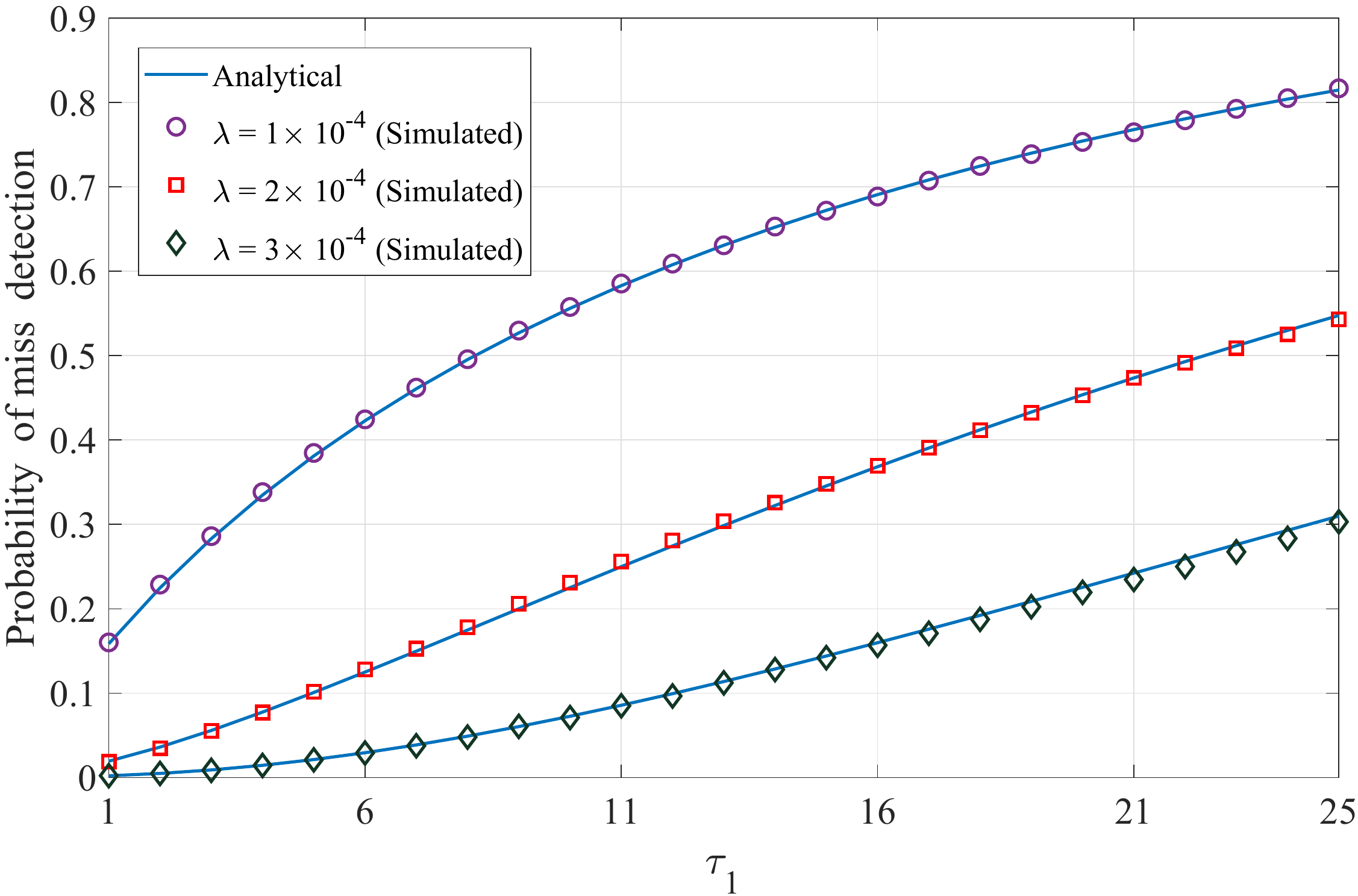}
	\end{center}
	\caption{Probability of miss detection versus threshold $\tau_1$. The probability of miss increases with an increase in the threshold value $\tau_1$ and decreases with increase in transmitter density.}
	\label{fig:4}
\end{figure}
\section{Numerical Results}

In this section, we present the numerical results for the probability of symbol error of the proposed system. All the derived analytical results are compared with the results obtained using Monte Carlo based simulation approach. For \eqref{eq:e2} particle based simulation is also used. In particle based simulation approach, the emitted molecules from the transmitter propagates via Brownian motion and are absorbed at the paired spherical absorbing receiver at the sink. The displacement of a single molecule in each dimension during the time $\Delta t$ is a normal distributed random variable with mean $0$ and variance $2D\Delta t$. The simulation step time $\Delta t$ is fixed as $10^{-4}$s and the receiver counts the absorbed molecules every $10^{-4}$s. In all simulations, the receiver radius $r_j=4\mu$m, symbol time $T_s=0.2$s, $\tau_0=0$ and $\tau_M=200$. In the $3$-D space, the location of the tagged receiver is fixed at the origin, and all the transmitter locations are generated outside the receiver volume uniformly and the simulations are carried out for $10^4$ realizations. For simulation, the PPP nodes are generated up to a distance ($R_a$) of $100\mu$m from the centre of the spherical receiver located at the origin. For a transmitter density of  $\lambda=1\times 10^{-4}$ per $\mu$m$^3$ there will be 418 transmitters on average outside receiver volume and inside the 3-D space of radius $100\mu$m. 

\subsection{System with One Source and One Sink}
 To analyze this system, we set $D=100\times 10^{-12}$m$^2/$s and $\mu_d=0 \ s^{-1}$ under the assumption that the molecules undergo diffusion without degradation. Fig. \ref{fig:3} shows the variation of the probability of symbol error at a spherical receiver with respect to the distance between the receiver surface and point transmitter. In Fig. \ref{fig:3}, the symbol error rate (SER) performance of OOK, BCSK and QCSK are compared using \eqref{eq:e1} and \eqref{eq:e2}. For OOK modulation, the number of molecules emitted by transmitter corresponding to symbol 0 and 1 is set as $Q_0=0$ and $Q_1=50$ with a central threshold value $ \tau_1=10$. In the case of BCSK, $Q_0=20$ and $Q_1=80$ number of molecules are emitted corresponding to symbol 0 and 1 with a central threshold values $ \tau_1=30$ molecules. For QCSK the number of molecules emitted corresponding to symbols 00, 01, 10 and 11 are   $Q_0=0$, $Q_1=20$, $Q_2=40$ and $Q_3=60$ with threshold values $\tau_1=10$, $\tau_2=20$ and $\tau_3=40$ for detection at the receiver. The previous bits are set as 01010101 and the probability of symbol error is calculated for the next symbol considering the ISI effect of all the previous symbols. It can be clearly observed in Fig. \ref{fig:3} that the symbol error rate is less for OOK since it emits no molecules corresponding to bit 0 and thus ISI is less. The QCSK has higher symbol error rate due to more molecules in medium which results ISI and also the gap between threshold levels is less.

\subsection{System with Multiple Sources and One Sink}
To analyze the system with multiple multiple sources, we further set $D=100\times 10^{-12}$m$^2/$s and $\mu_d=0 \ s^{-1}$. Fig. \ref{fig:4} shows the probability of miss detection (i.e., the probability of decoding $0$ conditioned on the fact that bit $1$ was sent) versus the detection threshold $\tau_1$. Without considering molecular degradation, $Q_1=50$ number of molecules are emitted from the paired transmitter and other interfering point transmitters  using OOK. Since all the transmitters are sending bit 1 and ISI is absent, with the increase in transmitter density the number of molecules reaching the receiver during that time instant increases. Thus the chance of net number of molecules crossing the threshold value $\tau_1$ increases and the error performance improves with  transmitter density. \

 Fig. \ref{fig:5} plot the symbol error probability of OOK and BCSK modulation with respect to decision threshold $\tau_1$ considering (i) without molecular degradation, i.e., $\mu_d=0\ s^{-1}$ and (ii) molecular degradation with $\mu_d=1 \ s^{-1}$. The transmitter density is $\lambda=1\times 10^{-4}$ per $\mu$m$^3$ and diffusion coefficient is $D=100\times10^{-11}$m$^2/$s. The simulation is done by considering the number of molecules transmitted by the point transmitter are $Q_0=0$ and $Q_1=50$ corresponding to bits 0 and 1 for OOK modulation and $Q_0=10$ and $Q_1=50$ corresponding to bits 0 and 1 for BCSK modulation. Using \eqref{eq:e1} and \eqref{eq:e8}, Fig. 5 is plotted by fixing previous symbol sequence as 010101010 and the probability of symbol error is calculated for arbitrary 10th symbol with channel memory length $L=5$. As a result of molecular degradation, the molecules contributing ISI and the molecules from interfering transmitters located far away degrades and the amount of ISI and MUI reduces. So molecular degradation improves the performance of the system which is evident from Fig. \ref{fig:5}. A careful trade-off between rate of molecular degradation and symbol time is required to avoid degradation of all molecules corresponding to current symbol before reaching the receiver.  Also, OOK modulation performs better than BCSK modulation since number of molecules resulting ISI is lesser for OOK than BCSK.\

 \begin{figure}
	\begin{center}
		\includegraphics[scale=0.41]{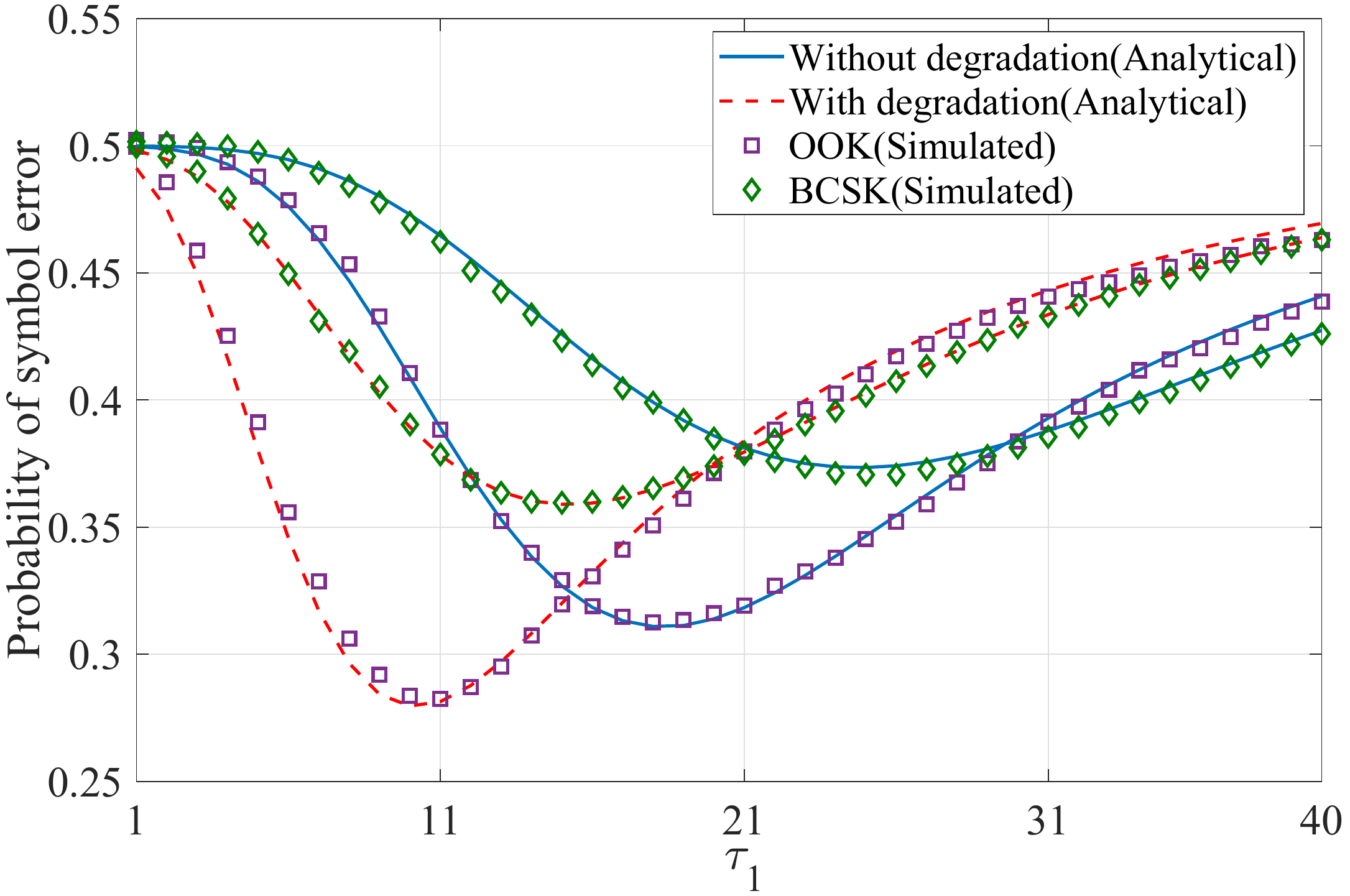}\
	\end{center}
	\caption{Probability of symbol error versus threshold $\tau_1$ for the detection of symbol $0$ and $1$. The symbol error reduces and then increases later with the increase in threshold value. Molecular degradation improves the system performance by reducing the symbol error probability.}
	\label{fig:5}
\end{figure}

\section{Conclusion}
\abc{In this paper, we proposed a new  hybrid modulation scheme for molecular communication  systems where each source and sink BNMs has multiple transmitters and receivers. We developed an analytical framework to analyze a system containing multiple sources and one sink.} The use of different molecule types in different transmitter-receiver links eliminates ILI and reduces MUI, which in turns improves the system performance.  Assuming the random position of sources as HPPP outside the receiver volume, the analytical expression for the probability of symbol error under ISI and MUI is derived for 3-D space using stochastic geometry. 
Results are compared for different $M$-ary CSK modulation schemes with and without considering the effect of MUI and it agrees with corresponding simulation results. Future works will focus on suitable detection schemes for the proposed system which can improve the performance.
\bibliographystyle{IEEEtran}
\bibliography{References}

\end{document}